\documentclass[%
 reprint,
 superscriptaddress,
 nofootinbib,
 amsmath,amssymb,
 aps,
pra,
longbibliography
]{revtex4-2}

\usepackage{graphicx}
\usepackage{dcolumn}
\usepackage{bm}
\usepackage{amsmath,amssymb}
\usepackage{mhchem}
\usepackage{braket}
\usepackage[dvipsnames]{xcolor}
\usepackage[
  colorlinks=true,
linkcolor=Blue,
  citecolor = Blue,
  urlcolor = Blue,
  unicode
  ]{hyperref}
\usepackage{natbib}
\usepackage[nameinlink]{cleveref}
\usepackage{float}
\crefname{equation}{Eq.}{Eqs.}
\Crefname{equation}{Equation}{Equations}
\crefname{figure}{Fig.}{Figs.}

\Crefname{figure}{Figure}{Figures}

\usepackage{ulem}
\usepackage{color}

\begin{document}

\preprint{APS/123-QED}

\title{Optimal Control of Coupled Sensor–Ancilla Qubits for Multiparameter Estimation}

\author{Ayumi Kanamoto} 
\thanks{These authors contributed equally.}
\affiliation{Department of Nuclear Science and Engineering, Massachusetts Institute of Technology, Cambridge, MA 02139, USA}
\affiliation{Department of Electrical and Electronic Engineering, Institute of Science Tokyo, Meguro,
Tokyo 152-8550, Japan}

\author{Takuya Isogawa} 
\thanks{These authors contributed equally.}
\affiliation{Department of Nuclear Science and Engineering, Massachusetts Institute of Technology, Cambridge, MA 02139, USA}

\author{Shunsuke Nishimura} 
\affiliation{Department of Nuclear Science and Engineering, Massachusetts Institute of Technology, Cambridge, MA 02139, USA}
\affiliation{Department of Physics, The University of Tokyo, Bunkyo-ku, Tokyo, 113-0033, Japan}

\author{Haidong Yuan} 
\affiliation{
   Department of Mechanical and Automation Engineering, The Hong Kong Institute of Quantum Information Science and Technology, State Key Laboratory of Quantum Information Technologies and Materials, The Chinese University of Hong Kong, Shatin, Hong Kong SAR, China}

\author{Paola Cappellaro} 
\email{pcappell@mit.edu}
\affiliation{Department of Nuclear Science and Engineering, Massachusetts Institute of Technology, Cambridge, MA 02139, USA}
\affiliation{Department of Physics, Massachusetts Institute of Technology, Cambridge, MA 02139, USA}

\begin{abstract}
Designing optimal control for multiparameter quantum sensing is essential for approaching the ultimate precision limits. However, analytical solutions are generally available only for simple systems, while realistic scenarios often involve coupled qubits and time-dependent Hamiltonians. Here we numerically investigate optimal control of a two-qubit sensor–ancilla system coupled via an Ising term using Gradient Ascent Pulse Engineering (GRAPE) to minimize the objective function. 
By seeding the optimization recursively with solutions obtained for smaller coupling strengths and selecting a suitable initial guess, we achieve robust convergence and high precision across a wide range of interaction strengths and field configurations.
The proposed approach offers a practical route toward high-sensitivity, robust multiparameter magnetometry and it is applicable to solid-state quantum sensors such as nitrogen-vacancy (NV) centers in realistic experimental settings.
\end{abstract}

\maketitle

\section{\label{sec:introduction}Introduction}

\begin{figure}
    \centering
    \includegraphics[width=0.9\columnwidth]
    {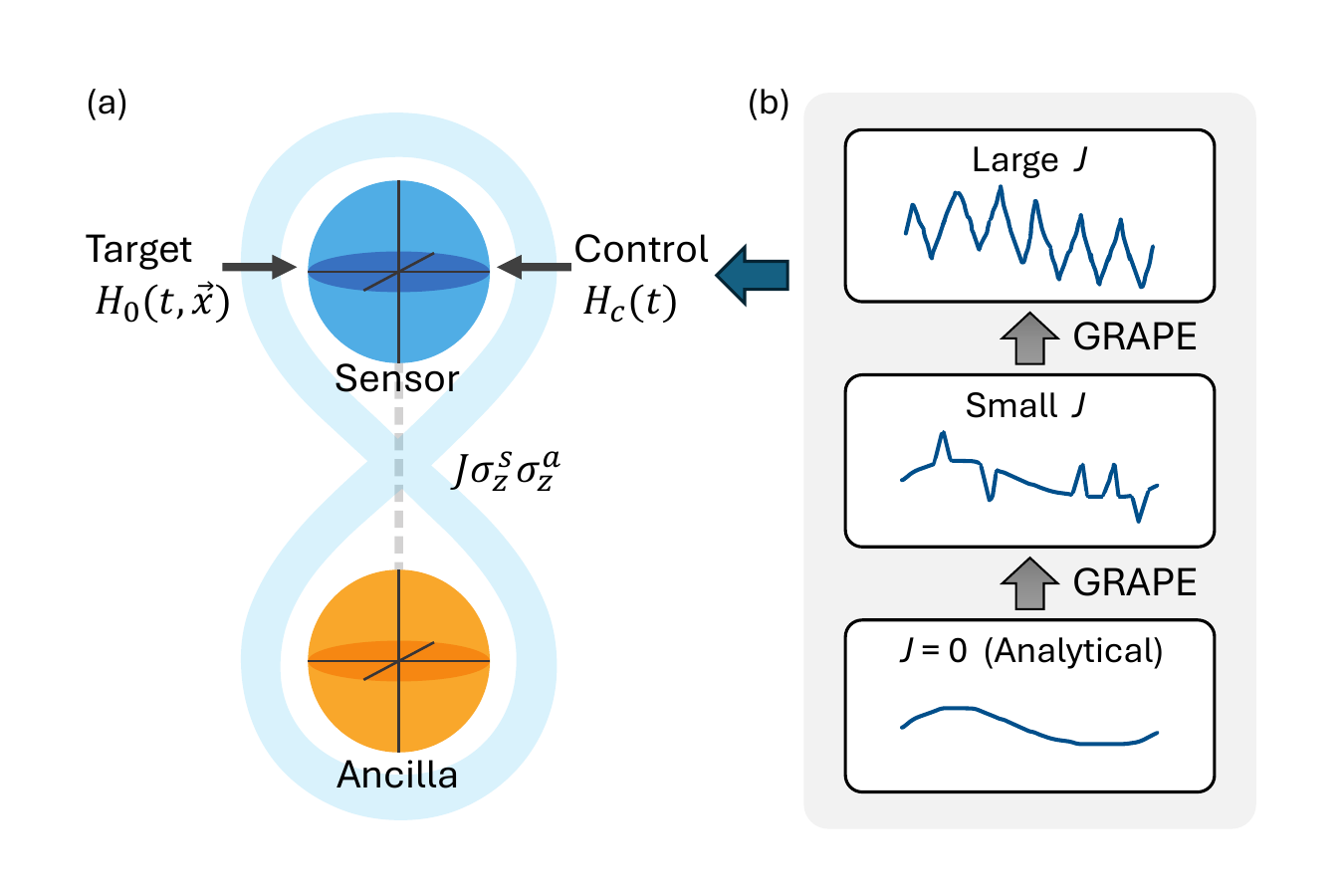}
    \caption{(a) Schematic of the two-qubit sensing system. A sensor qubit is coupled to an ancilla qubit via an Ising interaction $J\sigma_z^s\sigma_z^a$. The sensor is driven by a time-dependent control Hamiltonian $H_c(t)$ optimized with GRAPE while interacting with the target Hamiltonian $H_0(t,\vec{x})$ containing the unknown parameters $\vec{x}$.
(b) The recursive optimization strategy. To find optimal controls for systems with strong interactions (Large $J$), the algorithm bootstraps from the analytical solution for the non-interacting case ($J=0$). The GRAPE-optimized control for a small coupling $J$ serves as the initial guess for the subsequent optimization at a larger coupling, ensuring robust convergence.}
    \label{fig:schematic}
\end{figure}

Quantum sensing exploits quantum-mechanical resources to achieve better measurement precision~\cite{Giovannetti2011,RevModPhys.89.035002}. Introducing suitable control Hamiltonians can improve the sensitivity of quantum parameter estimation and a variety of protocols have been developed accordingly~\cite{Liu_2020, PhysRevLett.115.110401,PhysRevA.96.012117, pang2017optimal, PhysRevA.95.062342, PhysRevLett.123.250502, PhysRevLett.117.160801, PhysRevA.96.042114,xu2019generalizable, PhysRevA.103.042615,hu2024control}. In practical settings, several parameters, such as the vector components of the magnetic field and the amplitude and frequency of the AC field, are unknown at the same time. Their simultaneous estimation can surpass the precision attainable by estimating each parameter separately, providing an advantage in both sensitivity and resource efficiency. Such an advantage can be evaluated with the quantum Fisher information matrix (QFIM), which sets the ultimate bounds on achievable precision via the quantum Cram\'er-Rao bound~\cite{Liu_2020}. In multiparameter estimation, extensive research has been conducted on probe states that maximize the QFIM, as well as on optimal measurement strategies that achieve the corresponding bounds~\cite{PhysRevA.69.022303,Imai_2007,PhysRevLett.111.070403,PhysRevA.103.042615,hu2024control,PhysRevLett.125.020501,PhysRevA.94.052108,Liu_2020,vasilyev2024optimalmultiparametermetrologyquantum,PhysRevLett.117.160801,PhysRevLett.112.103604,Taylor2013,PhysRevLett.124.060502,Polino:19,Roccia_2018,Ciampini2016,Zhou:15,doi:10.1126/sciadv.abd2986,PhysRevLett.126.070503,PhysRevA.96.042114,xie2019optimal,isogawa2025entanglement}. Still, it is not always possible to simultaneously saturate the ultimate bounds for each parameter obtained in single-parameter estimation, as the optimal measurements or control strategies for different parameters are generally incompatible~\cite{Liu_2020,PhysRevA.94.052108,hu2024control}.

A promising strategy to address this issue is to design suitable control Hamiltonians through numerical optimization techniques, such as Gradient Ascent Pulse Engineering (GRAPE)~\cite{khaneja2005optimal}. GRAPE has been widely used in quantum control to design optimal (high-fidelity) pulse sequences in a variety of systems, including  NMR~\cite{khaneja2005optimal, tovsner2009optimal, de2011second}, nitrogen-vacancy (NV) centers~\cite{dolde2014high, rong2015experimental, PhysRevA.100.012110, liddy2023optimal}, and Bose-Einstein condensates~\cite{PhysRevA.90.033628, dionis2025optimal}. 
 More recently, GRAPE has also been applied to the optimization of control for quantum parameter estimation~\cite{PhysRevA.96.042114, PhysRevResearch.4.043057}. Some challenges still remain: tackling time-dependent target Hamiltonians is more complex, and only analytical approaches have been studied for sensing~\cite{PhysRevLett.117.160801,xie2019optimal,PhysRevLett.126.070503}. In addition, GRAPE performance depends on the quality of the initial guess for the control fields, as it is  susceptible to local optima whose number typically increases with the complexity of the Hamiltonian, such as the presence of interactions~\cite{Chakrabarti01102007}.

To tackle these challenges, here we investigate a two-qubit system consisting of a sensor and an ancilla qubit (Fig.~\ref{fig:schematic}(a)), which is a paradigmatic model in quantum multiparameter estimation~\cite{PhysRevA.69.022303,PhysRevLett.117.160801,PhysRevLett.126.070503,doi:10.1126/sciadv.abd2986,isogawa2025entanglement}. In particular, we consider an Ising interaction between the sensor and ancilla qubits and assume that control can be applied only to the sensor qubit. This setting is well aligned with many physical platforms, including NV centers, where the sensor electron spin and the ancilla nuclear spin are coupled via hyperfine interactions and engineering a single-spin Hamiltonian is straightforward~\cite{isogawa2025entanglement}. We apply GRAPE to optimize control pulses under both static and time-dependent target magnetic fields by introducing a bootstrapping protocol for the initial guess  (Fig.~\ref{fig:schematic}(b)), and we evaluate the resulting estimation precision. Notably, the sensitivity achieved by the GRAPE-optimized single-qubit control in the interacting case ($J>0$) approaches that of the interaction-free scenario ($J=0$), while still yielding sizable gains over naive control protocols that neglect the Ising coupling. 
Consequently, the framework is naturally compatible with NV center–based quantum sensors and holds potential for realistic solid-state quantum metrology.
 
\section{\label{sec:methods}Methods}
\subsection{Multiparameter Estimation}
We consider a qubit sensor coupled to an ancilla qubit via the Ising interaction 
\(J\sigma_z^s\sigma_z^a\), where \(J\) denotes the coupling strength, and $s,\,a$ label the sensor and ancilla qubits, respectively. This system could be embodied by an NV center electronic spin (the sensor) and its Nitrogen nuclear spin (the ancilla qubit) coupled by the hyperfine interaction~\cite{isogawa2025entanglement}. The goal is to measure the parameters of a Hamiltonian $H_0=\vec B(t)\cdot\vec\sigma+J\sigma_z^s\sigma_z^a$ acting on the qubit sensor, where $\vec B(t)$ contains multiple parameters $\vec{x} = (x_1, x_2, \ldots, x_d)$ of interest.

Performing a sensing protocol that prepares the (parameter-dependent) state $\rho_{\vec{x}}$ will yield measurement probabilities determined by a POVM $\{E(y)\}$ through $p_{y|\vec{x}} = \mathrm{Tr}[\rho_{\vec{x}} E(y)]$. The classical Fisher information matrix (CFIM) obtained from these probabilities quantifies the attainable precision via the classical Cram\'er–Rao bound with the specific POVM measurement. Optimizing over POVMs yields the QFIM, which provides a measurement-independent precision limit. 

The optimal protocol for this qubit-ancilla system has been found to require a probe state  prepared in the maximally entangled Bell state \(\ket{\Phi^+} = ( \ket{00} + \ket{11} )/\sqrt{2}\)~\cite{PhysRevLett.117.160801}. As we are interested in finding the optimal control, we keep the measurement fixed to a Bell measurement in a basis rotated by an angle of \(\pi/3\)~\cite{isogawa2025entanglement} and evaluate the performance using the optimization objective function constructed from the diagonal elements of the CFIM since in multiparameter estimation, the QFIM bound generally does not guarantee unattainable precision due to the incompatibility of optimal measurements~\cite{Liu_2020,PhysRevA.94.052108}. 
For this reason, throughout this work, we adopt the strategy of keeping the probe state and measurement basis fixed and optimizing only the control Hamiltonian.

\subsection{GRAPE Algorithm}
We use the GRAPE algorithm~\cite{PhysRevA.96.042114} to find the optimal single-qubit control Hamiltonian $H_c(t)$ for the multiparameter estimation task, applied simultaneously to the target Hamiltonian $H_0(t)$. We note that while analytical approaches suggest a two-qubit control Hamiltonian designed to cancel the Ising interaction \(J\sigma_z^s\sigma_z^a\), such schemes are typically complex to implement. For example, rather than simultaneous control, where the target field and the control field are applied concurrently, they can be realized through sequential control based on the Trotter formula~\cite{isogawa2025entanglement}. This, however, requires stronger assumptions on the target field as well as greater time overhead. Thus, we assume the two qubits evolve unitarily under the total Hamiltonian

\begin{equation}
    H(t) = H_0(t) + H_c(t) = H_0(t) +
    \sum_{k \in \{x, y, z\}} V_k(t)\, H_k
\end{equation}
where
\(H_k = \sigma_{k}^s=\sigma_k^s\otimes I^a\).
Here, \(V_k(t)\) represents the time-dependent control amplitude acting along the \(k\)-direction on the sensor qubit and is implemented as a piecewise-constant function, where each segment corresponds to a digital time step determined by the experimental setup. Accordingly, we divide the total evolution time \(T\) into \(N\) discrete steps, with each time step of width \(\tau = T/N\). At each step, the state is updated using the unitary operator \(U_i = \exp(-i H(t_i) \tau)\), allowing us to approximate the time-evolved density matrix as:
\begin{equation}
    \rho(T) = U_N \cdots U_2 U_1 \, \rho(0) \, U_1^\dagger U_2^\dagger \cdots U_N^\dagger.
\end{equation}

We employ the GRAPE algorithm to obtain the optimal control fields that maximize the objective function \[f_0(T)=\left( \sum_{\alpha} \frac{1}{\mathcal{F}_{\mathrm{cl},\alpha\alpha}(T)} \right)^{-1},\] 
which provides a compact measure of the overall estimation precision~\cite{PhysRevA.96.042114}.

To improve the performance of the GRAPE optimization, we introduce a recursive protocol, where we first find a solution for a smallest value $J$  using the known solution for $J=0$ as the initial guess. We then use the found solution as an initial guess for a larger $J$. In this way, we can robustly  find optimal solutions for large $J$ without the algorithm being trapped in local (low-performing) minima. 
The functional gradient of the objective function, \(\delta f_0(T) / \delta V_k(t)\), was evaluated using finite-difference numerical differentiation because this method is simple and practical to implement. In the implementation, computational acceleration was achieved by employing Numba-based JIT compilation and by improving the efficiency of matrix exponential calculations.
The source code used in this study is available online~\cite{githubgrape}.

\begin{figure}[bt]
    \centering
    \includegraphics[width=0.95\columnwidth]
    {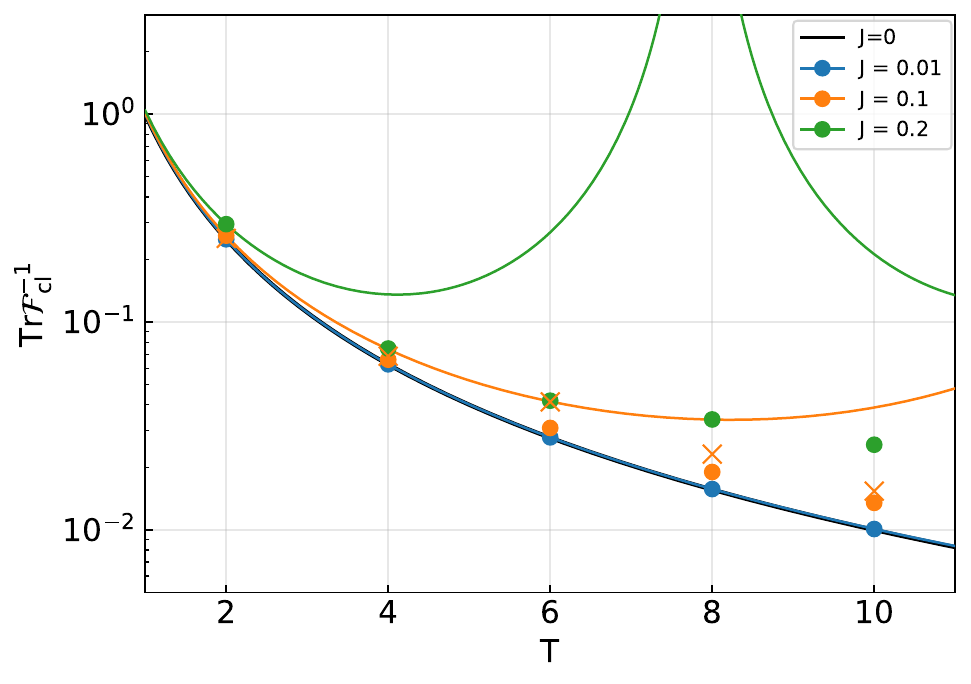}
    \caption{Estimation precision \(\mathrm{Tr}[F_{\mathrm{cl}}^{-1}]\) as a function of total evolution time \(T\) for different values of the Ising interaction strength \(J\) for Eq.~\ref{eq:SMF}. The dots indicate results obtained using GRAPE with Eq.~\ref{eq:SMF_ctrl} as the initial guess, while the crosses denote those obtained with randomly generated initial guesses. The solid lines indicate the precision limits when Eq.~\ref{eq:SMF_ctrl} is applied to Eq.~\ref{eq:SMF}. The black, blue, orange, and green colors correspond to \(J=0\), 0.01, 0.1, and 0.2 respectively. The parameters are fixed as \(B = 1\), \(\theta = \pi/4\), and \(\phi = \pi/4\).}
    \label{fig:ising vector sensing}
\end{figure}

\begin{figure*}
    \centering
    \includegraphics[width=0.9\textwidth]{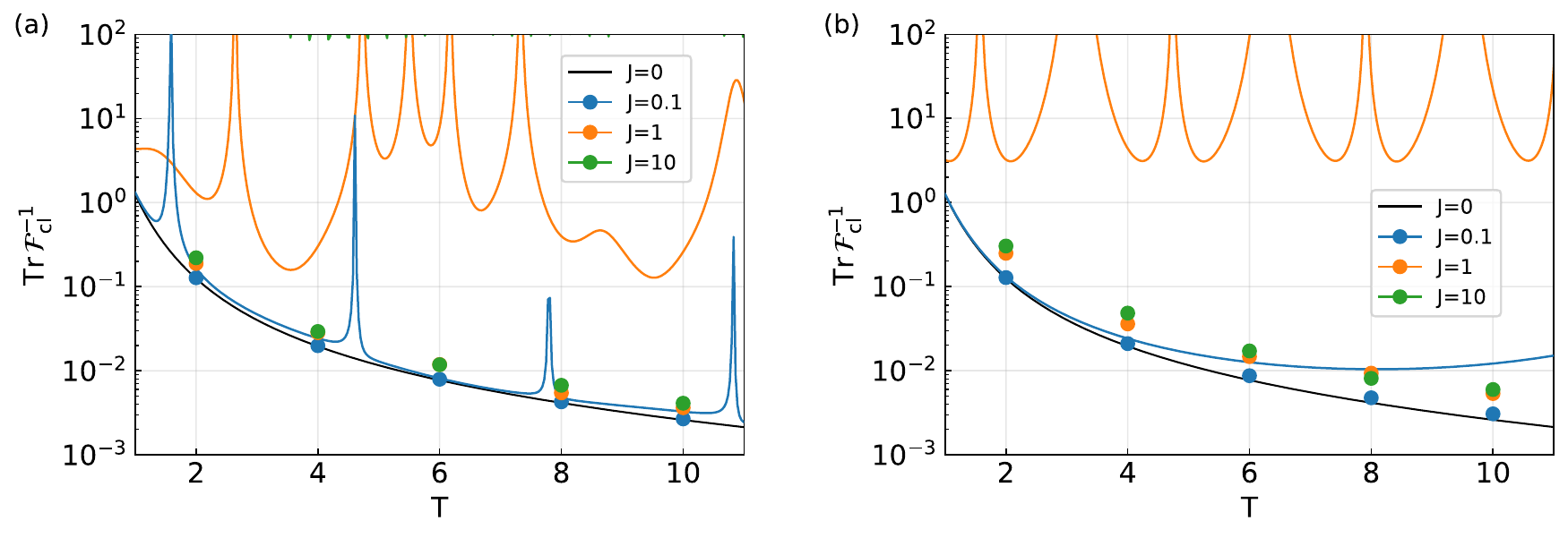}
    \caption{Estimation precision \(\mathrm{Tr}[F_{\mathrm{cl}}^{-1}]\) as a function of total evolution time \(T\) for different values of the Ising interaction strength \(J\) for  Eq.~\ref{eq:CPL_xz} with (a) an xz-plane field 
    and (b) an xy-plane field.
    The dots represent the results obtained by GRAPE and the solid lines indicate the precision limits when the optimal control designed for the Hamiltonian without Ising terms is applied to the Hamiltonian with an Ising term. The black, blue, orange, and green correspond to \(J=0\), 0.1, 1, and 10 respectively. The y-axis is capped at \(10^2\) for better visibility and most of the green curve is clipped because its floor is \(10^2\) in (a) and \(10^4\) in (b). The parameters are fixed as \(B = \omega=1\).}
    \label{fig:CPL}
\end{figure*}

\section{\label{sec:Results}Results}
To demonstrate the potential of our GRAPE optimization protocol we consider four target Hamiltonians, all including the Ising interaction, that range from more simple models to more complex, but practically relevant scenarios.
First we consider estimating a static magnetic field as a baseline optimization task; indeed, the corresponding time-independent Hamiltonian reduces computational cost and allows efficient analysis of GRAPE's performance.
We then consider time-dependent target Hamiltonians. We start by estimating a circularly polarized field in the xz-plane and the xy-plane, for which an optimal control for the non-interacting case $J=0$ is known. This allows us to start with a good initial guess and compare our numerical results to the $J=0$ precision limit. We finally turn to estimation tasks that are more practical, namely a linearly polarized field.

\subsection{\label{sec:level3.1}Static magnetic field}
We consider the target Hamiltonian
\begin{equation}
    H_0 = \vec{B} \cdot \vec{\sigma}^s + J \sigma_z^s \sigma_z^a
    \label{eq:SMF}
\end{equation}
with the static magnetic field parameterized as \(\vec{B}=(B \sin\theta \cos\phi, B \sin\theta \sin\phi, B\cos\theta)\), and focus on the simultaneous estimation of the three parameters \(B\), \(\theta\), and \(\phi\). When the target Hamiltonian is time-independent and  \(J=0\), the optimal control is given by~\cite{pang2017optimal, PhysRevLett.117.160801}
\begin{equation}
    H_c=-H_0.
    \label{eq:SMF_ctrl}
\end{equation} 
For the Hamiltonian in Eq.~\ref{eq:SMF}, the control enables the estimation protocol to saturate the precision limit, yielding \(\mathrm{Tr}[F_{\mathrm{cl}}^{-1}] = \frac{1}{4T^2} \left( 1 + \frac{1}{B^2} + \frac{1}{B^2 \sin^2\theta} \right)\)~\cite{PhysRevLett.117.160801}. In Figure~\ref{fig:ising vector sensing}, we set \(B=1\) and \(\theta=\phi=\pi/4\), for which the precision bound simplifies to \(\mathrm{Tr}[F_{\mathrm{cl}}^{-1} ]= \frac{1}{T^2}\) (black line).
Optimal single-qubit control for $J>0$ is not known, thus we applied GRAPE sequentially for \(J=0.01, 0.1, 0.15,\) and \(0.2\), using Eq.~\ref{eq:SMF_ctrl}  as the initial guess for \(J=0.01\) and the optimal control obtained by GRAPE in the immediately preceding case for the other $J$ values. Figure~\ref{fig:ising vector sensing} compares the resulting performance (dots) as a function of the evolution time $T$ with that obtained when Eq.~\ref{eq:SMF_ctrl} was directly applied  (solid lines), demonstrating the improvement in estimation precision obtained by GRAPE.
Furthermore, while the application of Eq.~\ref{eq:SMF_ctrl} gives rise to points where the precision periodically reaches local maxima (\(T=n\pi/2J\)), the use of GRAPE was found to achieve robust control even at such points. For example, the plot for $J = 0.2$ at $T = 8$ lies closest to a peak and $\mathrm{Tr}[F_{\mathrm{cl}}^{-1}]$ is reduced by $99.9\%$ from $35.2$ to $0.034$.

Figure~\ref{fig:ising vector sensing} also compares the estimation precision when the initial guess for \(J=0.01\) was set to Eq.~\ref{eq:SMF_ctrl} (dots) and when it was set to randomly generated values (crosses). In both cases, for the larger values of \(J\), the initial guess for GRAPE was taken as the optimized control obtained at the immediately preceding \(J\). The former consistently yielded higher precision, with the difference reaching up to approximately \(25\%\). This result reflects the fact that GRAPE is a heuristic algorithm that depends on the choice of the initial value, and demonstrates that the optimal control for the Hamiltonian without the Ising term serves as a better initial guess than randomly generated controls.

\subsection{\label{sec:level3.2}Circularly polarized field}
We consider the estimation of two parameters, the amplitude \(B\) and the frequency \(\omega\) corresponding to a circularly polarized field in the x$\alpha$-plane, with $\alpha=$ z or y. The target Hamiltonian including the Ising interaction, is then 
\begin{equation}
    H_0(t) = -B \left[ \cos(\omega t)\, \sigma_x^s + \sin(\omega t)\, \sigma_\alpha^s \right] + J\, \sigma_z^s \sigma_z^a. 
    \label{eq:CPL_xz}
\end{equation}

In the absence of the Ising term, the optimal single-qubit control was found to be~\cite{xie2019optimal, PhysRevLett.126.070503} 
\begin{equation}
\begin{split}
    H_c(t) &= -H_0(t) +i \frac{\omega}{4} [\sigma^s_\alpha,\sigma^s_x], 
           \label{eq:CPL_xz_ctrl}
\end{split}
\end{equation}
which saturates the fundamental limit of the attainable precision, \(\mathrm{Tr}[F_{\mathrm{cl}}^{-1}] =\Delta^2B+\Delta^2\omega=\frac{1}{4T^2}+\frac{1}{B^2T^4}\).

The Ising term however breaks the symmetry between the y and z directions, and the two cases need to be considered separately. 

In particular, in the case of the xz-plane field,
the transformation to a frame rotating around the y-axis imposes a time dependence on the Ising term that partially refocuses it, enabling the treatment of larger \(J\) values than in the static magnetic field case. Conversely, the rotating frame transformation for the xy-plane field yields a time independent Hamiltonian; still, the presence of a (potentially large) term $\propto \omega$ mitigates the effects of the Ising term.
Thus we applied GRAPE to Eq.~\ref{eq:CPL_xz}
for \(J=0.01, 0.1, 1, 5,\) and \(10\). For \(J=0.01\), Eq.~\ref{eq:CPL_xz_ctrl} 
was used as the initial guess, whereas for larger \(J\) values the optimized control obtained at the previous smaller \(J\) with the same evolution time $T$ was employed as the initial guess.  Figure~\ref{fig:CPL} compares the estimation precision obtained using GRAPE-optimized controls (dots) with that obtained by directly applying Eq.~\ref{eq:CPL_xz_ctrl}  (solid lines) for \(J=0.1, 1,\) and \(10\). Figure~\ref{fig:CPL}(a) corresponds to the xz-plane case, while Fig.~\ref{fig:CPL}(b) corresponds to the xy-plane case. In both cases, GRAPE consistently outperforms the direct application of Eq.~\ref{eq:CPL_xz_ctrl}. As shown in Fig.~\ref{fig:CPL}(a), applying Eq.~\ref{eq:CPL_xz_ctrl} to Eq.~\ref{eq:CPL_xz} results in non-periodic peaks in the xz-plane field, while it  produces periodic peaks for the xy-plane field as shown in Fig.~\ref{fig:CPL}(b). In contrast, GRAPE achieves robust control approaching the precision limit at all times $T$. In particular,  for \(J=10\)  \(\mathrm{Tr}[F_{\mathrm{cl}}^{-1}]\) is reduced by six orders of magnitude (from $3.56\times10^{6}$ to $6.71\times10^{-3}$) at \(T=8\) for the xz case and by eight orders of magnitude (from $3.01\times10^{6}$ to $1.72\times10^{-2}$) at \(T=6\) in the xy case.

\subsection{\label{sec:level3.3}Linearly polarized field}
Next we consider the simultaneous estimation of two parameters, the amplitude \(B\) and the frequency \(\omega\), under the Hamiltonian describing a linearly polarized field with an Ising interaction:
\begin{equation}
    H_0(t) = B \cos(\omega t)\, \sigma_x^s + J\, \sigma_z^s \sigma_z^a.
    \label{eq:LPL}
\end{equation}
When \(J = 0\), an optimal single-qubit control Hamiltonian
\begin{equation}
    H_c(t) = -H_0(t) + \frac{\omega}{2} \sigma_y^s,
    \label{eq:LPL_ctrl}
\end{equation}
is obtained in the long time limit \(T \gg 2\pi/\omega\), and the precision bound \(\mathrm{Tr}[F_{\mathrm{cl}}^{-1}]=1/T^2+4/(B^2 T^4)\) is achieved~\cite{multiparam_AC}. We note that we could equivalently apply a rotating frame around $\sigma_z$. Since in the circularly polarized case we obtained slightly better performance for the xz-case than the xy case, we choose to use the $\sigma_y$ frame. This results in a time-dependent Hamiltonian when including the Ising term. As the frequency $\omega$ that satisfies the condition \(T \gg 2\pi/\omega\) needs to be large, this requires a correspondingly large number of time slices, making the computation prohibitively time-consuming. We thus investigated  a shorter evolution time \(T\) than in the static and circularly polarized field cases.

\begin{figure}[b]
    \centering
    \includegraphics[width=8cm]{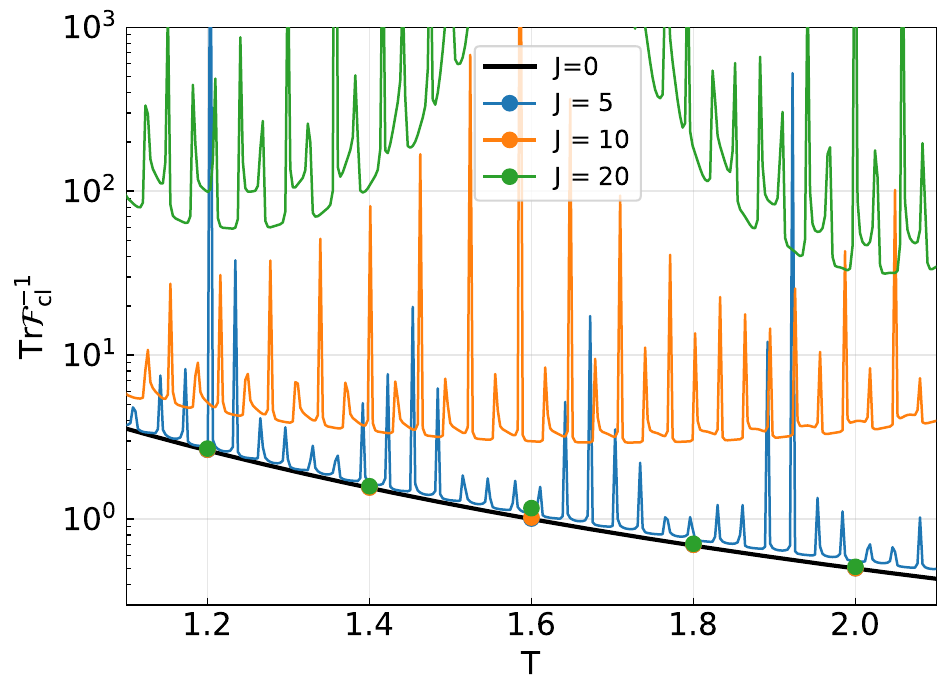}
    \caption{Estimation precision \(\mathrm{Tr}[F_{\mathrm{cl}}^{-1}]\) as a function of total evolution time \(T\) for different values of the Ising interaction strength \(J\) for Eq.~\ref{eq:LPL}. The dots represent the results obtained by GRAPE and the solid lines indicate the precision limits when Eq.~\ref{eq:LPL_ctrl} is applied to Eq.~\ref{eq:LPL}. The black, blue, orange, and green colors correspond to \(J=0\), 5, 10, and 20 respectively. The parameters are fixed as \(B = 1\) and \(\omega=100\).}
    \label{fig:LPL}
\end{figure}

\begin{figure*}
    \centering
    \includegraphics[width=0.9\textwidth]{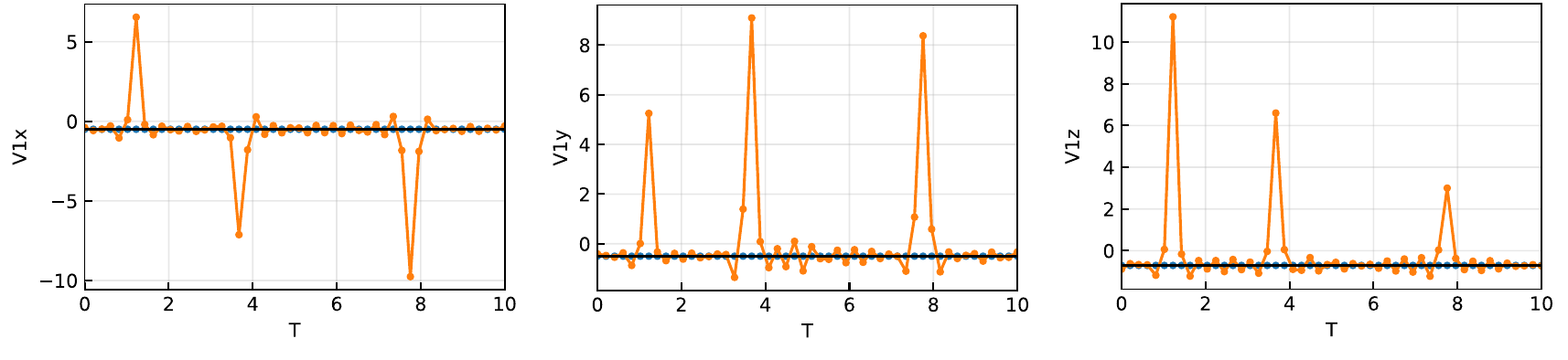}
    \caption{Comparison of the GRAPE-optimized controls for the static magnetic field Hamiltonian with an Ising interaction at \(J=0.01\) and \(T=10\) along the (a) x-, (b) y-, and (c) z-directions. The blue curve corresponds to optimization initialized with the optimal control for the static magnetic field without the Ising interaction, while the orange curve corresponds to optimization initialized with randomly generated controls. The black solid line denotes the optimal control for the static magnetic field without the Ising interaction.}
    \label{fig:SMF_controls_comparison}
\end{figure*}
Because the Ising interaction is effectively averaged out at large \(\omega\) and the dynamics approach the $J \approx 0$ limit, reasonably good precision is already achieved even before GRAPE optimization, which allows us to investigate larger values of \(J\) than in the other cases. We applied GRAPE to Eq.~\ref{eq:LPL} for six values of the Ising interaction strength, \(J = 0.01, 0.1, 1, 5, 10,\) and \(20\). For \(J = 0.01\), Eq.~\ref{eq:LPL_ctrl} was used as the initial guess, while for larger values of \(J\), the optimized control obtained at the next smaller \(J\) with the same evolution time \(T\) was adopted as the initial guess. The results for \(J = 5, 10,\) and \(20\) are shown as dots in Fig.~\ref{fig:LPL}, while the direct application of Eq.~\ref{eq:LPL_ctrl} without optimization is shown as solid lines. Compared with the solid lines, the GRAPE-optimized controls stably achieve estimation precision close to the fundamental limit. In particular, for \(J = 20\) and \(T = 1.6\), \(\mathrm{Tr}[F_{\mathrm{cl}}^{-1}]\) is reduced from \(2.92\times10^{5}\) to \(1.17\), corresponding to an improvement of five orders of magnitude.

\section{\label{sec:level4}Conclusion}
Designing optimal controls for multiparameter quantum sensing is challenging and intuitive analytical solutions can only be found for the simplest scenarios. 
Here we have shown how gradient-based optimization methods can find control protocols that closely approach the fundamental precision limits of multiparameter sensing. We made the control optimization  more robust by introducing a strategy that uses known analytical solutions for simple Hamiltonians to start a search for more complex interaction scenarios, focusing in particular on tackling the existence of Ising-type sensor–ancilla interactions and time-dependence. A simple recursive seeding strategy, which bootstraps solutions from the non-interacting case to progressively larger couplings, proved crucial for avoiding poor local optima and yielded robust controls over a broad range of evolution times.  Our bootstrapping technique could be further complemented by reinforcement learning (RL) techniques that could enable more adaptive and global exploration of the control landscape~\cite{xu2019generalizable, long2025optimal}.
Beyond the improvements achieved in our numerical benchmarks, the framework maps naturally onto solid-state platforms such as NV-center sensors coupled to nuclear ancillae, where control is digital, interactions are unavoidable, and multiparameter estimation arises naturally from vectorial magnetic fields or AC driving~\cite{isogawa2025entanglement}. Our control approach is compatible with available hardware using piecewise-constant controls, it can handle time dependence and coupling, and thus it can be immediately applied using NV centers in diamond or similar systems to achieve high-precision, robust quantum magnetometry in realistic experimental settings.

\appendix
\section{Controls obtained by GRAPE}
Figure~\ref{fig:SMF_controls_comparison} compares the controls obtained via GRAPE when the initial guess for the smallest $J$ is the optimal control for the static magnetic field without the Ising term and when it is given by randomly generated values. In both cases, the values of $\mathrm{Tr}[F_{\mathrm{cl}}^{-1}]$ are nearly identical to the precision limit; however, the resulting control waveforms differ significantly. The former yields a waveform almost identical to the optimal control for the static magnetic field without the Ising term, whereas the latter exhibits several large peaks during the evolution. When random initial guesses are employed, such waveforms can arise that are difficult to implement experimentally. This indicates that, even if high precision is achieved, the control may not necessarily be practical, thereby highlighting the importance of the initial guess in GRAPE.

\begin{acknowledgments}
This study was carried out using the TSUBAME4.0 supercomputer at Institute of Science Tokyo. This work was in part supported by NSF via grant (PHY CUA, Qusec). T.~Isogawa acknowledges support from a Mathworks fellowship. 
\end{acknowledgments}

\bibliography{ref}

@article{Chakrabarti01102007,
author = {Raj Chakrabarti and Herschel Rabitz},
title = {Quantum control landscapes},
journal = {International Reviews in Physical Chemistry},
volume = {26},
number = {4},
pages = {671--735},
year = {2007},
publisher = {Taylor \& Francis},
doi = {10.1080/01442350701633300},


URL = { 
    
        https://doi.org/10.1080/01442350701633300
    
    

},
eprint = { 
    
        https://doi.org/10.1080/01442350701633300
    
    

}

}

@article{PhysRevA.94.052108,
  title = {Compatibility in multiparameter quantum metrology},
  author = {Ragy, Sammy and Jarzyna, Marcin and Demkowicz-Dobrza\ifmmode \acute{n}\else \'{n}\fi{}ski, Rafa\l{}},
  journal = {Phys. Rev. A},
  volume = {94},
  issue = {5},
  pages = {052108},
  numpages = {11},
  year = {2016},
  month = {Nov},
  publisher = {American Physical Society},
  doi = {10.1103/PhysRevA.94.052108},
  url = {https://link.aps.org/doi/10.1103/PhysRevA.94.052108}
}

@article{
doi:10.1126/sciadv.abd2986,
author = {Zhibo Hou  and Jun-Feng Tang  and Hongzhen Chen  and Haidong Yuan  and Guo-Yong Xiang  and Chuan-Feng Li  and Guang-Can Guo },
title = {Zero\&\#x2013;trade-off multiparameter quantum estimation via simultaneously saturating multiple Heisenberg uncertainty relations},
journal = {Science Advances},
volume = {7},
number = {1},
pages = {eabd2986},
year = {2021},
doi = {10.1126/sciadv.abd2986},
URL = {https://www.science.org/doi/abs/10.1126/sciadv.abd2986},
eprint = {https://www.science.org/doi/pdf/10.1126/sciadv.abd2986}}

@article{Zhou:15,
author = {Xiao-Qi Zhou and Hugo Cable and Rebecca Whittaker and Peter Shadbolt and Jeremy L. O'Brien and Jonathan C. F. Matthews},
journal = {Optica},
number = {6},
pages = {510--516},
publisher = {Optica Publishing Group},
title = {Quantum-enhanced tomography of unitary processes},
volume = {2},
month = {Jun},
year = {2015},
url = {https://opg.optica.org/optica/abstract.cfm?URI=optica-2-6-510},
doi = {10.1364/OPTICA.2.000510},
}

@Article{Ciampini2016,
author={Ciampini, Mario A.
and Spagnolo, Nicol{\`o}
and Vitelli, Chiara
and Pezz{\`e}, Luca
and Smerzi, Augusto
and Sciarrino, Fabio},
title={Quantum-enhanced multiparameter estimation in multiarm interferometers},
journal={Scientific Reports},
year={2016},
month={Jul},
day={06},
volume={6},
number={1},
pages={28881},
issn={2045-2322},
doi={10.1038/srep28881},
url={https://doi.org/10.1038/srep28881}
}

@article{Roccia_2018,
doi = {10.1088/2058-9565/aa9212},
url = {https://dx.doi.org/10.1088/2058-9565/aa9212},
year = {2017},
month = {oct},
publisher = {IOP Publishing},
volume = {3},
number = {1},
pages = {01LT01},
author = {Roccia, Emanuele and Gianani, Ilaria and Mancino, Luca and Sbroscia, Marco and Somma, Fabrizia and Genoni, Marco G and Barbieri, Marco},
title = {Entangling measurements for multiparameter estimation with two qubits},
journal = {Quantum Science and Technology}
}

@article{Polino:19,
author = {Emanuele Polino and Martina Riva and Mauro Valeri and Raffaele Silvestri and Giacomo Corrielli and Andrea Crespi and Nicol\`{o} Spagnolo and Roberto Osellame and Fabio Sciarrino},
journal = {Optica},
number = {3},
pages = {288--295},
publisher = {Optica Publishing Group},
title = {Experimental multiphase estimation on a chip},
volume = {6},
month = {Mar},
year = {2019},
url = {https://opg.optica.org/optica/abstract.cfm?URI=optica-6-3-288},
doi = {10.1364/OPTICA.6.000288},
}

@article{PhysRevLett.124.060502,
  title = {Experimental Optimal Orienteering via Parallel and Antiparallel Spins},
  author = {Tang, Jun-Feng and Hou, Zhibo and Shang, Jiangwei and Zhu, Huangjun and Xiang, Guo-Yong and Li, Chuan-Feng and Guo, Guang-Can},
  journal = {Phys. Rev. Lett.},
  volume = {124},
  issue = {6},
  pages = {060502},
  numpages = {6},
  year = {2020},
  month = {Feb},
  publisher = {American Physical Society},
  doi = {10.1103/PhysRevLett.124.060502},
  url = {https://link.aps.org/doi/10.1103/PhysRevLett.124.060502}
}

@Article{Taylor2013,
author={Taylor, Michael A.
and Janousek, Jiri
and Daria, Vincent
and Knittel, Joachim
and Hage, Boris
and Bachor, Hans-A.
and Bowen, Warwick P.},
title={Biological measurement beyond the quantum limit},
journal={Nature Photonics},
year={2013},
month={Mar},
day={01},
volume={7},
number={3},
pages={229-233},
issn={1749-4893},
doi={10.1038/nphoton.2012.346},
url={https://doi.org/10.1038/nphoton.2012.346}
}

@article{PhysRevLett.112.103604,
  title = {Supersensitive Polarization Microscopy Using NOON States of Light},
  author = {Israel, Yonatan and Rosen, Shamir and Silberberg, Yaron},
  journal = {Phys. Rev. Lett.},
  volume = {112},
  issue = {10},
  pages = {103604},
  numpages = {4},
  year = {2014},
  month = {Mar},
  publisher = {American Physical Society},
  doi = {10.1103/PhysRevLett.112.103604},
  url = {https://link.aps.org/doi/10.1103/PhysRevLett.112.103604}
}

@misc{vasilyev2024optimalmultiparametermetrologyquantum,
      title={Optimal Multiparameter Metrology: The Quantum Compass Solution}, 
      author={Denis V. Vasilyev and Athreya Shankar and Raphael Kaubruegger and Peter Zoller},
      year={2024},
      eprint={2404.14194},
      archivePrefix={arXiv},
      primaryClass={quant-ph},
      url={https://arxiv.org/abs/2404.14194}, 
}

@article{PhysRevLett.125.020501,
  title = {Minimal Tradeoff and Ultimate Precision Limit of Multiparameter Quantum Magnetometry under the Parallel Scheme},
  author = {Hou, Zhibo and Zhang, Zhao and Xiang, Guo-Yong and Li, Chuan-Feng and Guo, Guang-Can and Chen, Hongzhen and Liu, Liqiang and Yuan, Haidong},
  journal = {Phys. Rev. Lett.},
  volume = {125},
  issue = {2},
  pages = {020501},
  numpages = {6},
  year = {2020},
  month = {Jul},
  publisher = {American Physical Society},
  doi = {10.1103/PhysRevLett.125.020501},
  url = {https://link.aps.org/doi/10.1103/PhysRevLett.125.020501}
}

@article{Imai_2007,
doi = {10.1088/1751-8113/40/16/009},
url = {https://dx.doi.org/10.1088/1751-8113/40/16/009},
year = {2007},
month = {mar},
publisher = {},
volume = {40},
number = {16},
pages = {4391},
author = {Imai, Hiroshi and Fujiwara, Akio},
title = {Geometry of optimal estimation scheme for SU(D) channels},
journal = {Journal of Physics A: Mathematical and Theoretical}
}

@article{PhysRevA.69.022303,
  title = {Estimation of unitary quantum operations},
  author = {Ballester, Manuel A.},
  journal = {Phys. Rev. A},
  volume = {69},
  issue = {2},
  pages = {022303},
  numpages = {6},
  year = {2004},
  month = {Feb},
  publisher = {American Physical Society},
  doi = {10.1103/PhysRevA.69.022303},
  url = {https://link.aps.org/doi/10.1103/PhysRevA.69.022303}
}

@article{RevModPhys.89.035002,
  title = {Quantum sensing},
  author = {Degen, C. L. and Reinhard, F. and Cappellaro, P.},
  journal = {Rev. Mod. Phys.},
  volume = {89},
  issue = {3},
  pages = {035002},
  numpages = {39},
  year = {2017},
  month = {Jul},
  publisher = {American Physical Society},
  doi = {10.1103/RevModPhys.89.035002},
  url = {https://link.aps.org/doi/10.1103/RevModPhys.89.035002}
}

@Article{Giovannetti2011,
author={Giovannetti, Vittorio
and Lloyd, Seth
and Maccone, Lorenzo},
title={Advances in quantum metrology},
journal={Nature Photonics},
year={2011},
month={Apr},
day={01},
volume={5},
number={4},
pages={222-229},
issn={1749-4893},
doi={10.1038/nphoton.2011.35},
url={https://doi.org/10.1038/nphoton.2011.35}
}

@article{multiparam_AC,
  title = {Approaching the Limit in Multiparameter AC Magnetometry with Quantum Control},
author = {Isogawa, Takuya and others},
journal = {To appear on arXiv},
}

@article{PhysRevLett.126.070503,
  title = {``Super-Heisenberg'' and Heisenberg Scalings Achieved Simultaneously in the Estimation of a Rotating Field},
  author = {Hou, Zhibo and Jin, Yan and Chen, Hongzhen and Tang, Jun-Feng and Huang, Chang-Jiang and Yuan, Haidong and Xiang, Guo-Yong and Li, Chuan-Feng and Guo, Guang-Can},
  journal = {Phys. Rev. Lett.},
  volume = {126},
  issue = {7},
  pages = {070503},
  numpages = {6},
  year = {2021},
  month = {Feb},
  publisher = {American Physical Society},
  doi = {10.1103/PhysRevLett.126.070503},
  url = {https://link.aps.org/doi/10.1103/PhysRevLett.126.070503}
}

@article{Liu_2020,
doi = {10.1088/1751-8121/ab5d4d},
url = {https://dx.doi.org/10.1088/1751-8121/ab5d4d},
year = {2019},
month = {dec},
publisher = {IOP Publishing},
volume = {53},
number = {2},
pages = {023001},
author = {Jing Liu and Haidong Yuan and Xiao-Ming Lu and Xiaoguang Wang},
title = {Quantum Fisher information matrix and multiparameter estimation},
journal = {Journal of Physics A: Mathematical and Theoretical}
}

@misc{githubgrape,
  title = {Code and results are available at \url{https://github.com/Daaolung/multiparam\_GRAPE}}
}

@article{PhysRevA.96.042114,
  title = {Control-enhanced multiparameter quantum estimation},
  author = {Liu, Jing and Yuan, Haidong},
  journal = {Phys. Rev. A},
  volume = {96},
  issue = {4},
  pages = {042114},
  numpages = {11},
  year = {2017},
  month = {Oct},
  publisher = {American Physical Society},
  doi = {10.1103/PhysRevA.96.042114},
  url = {https://link.aps.org/doi/10.1103/PhysRevA.96.042114}
}

@article{pang2017optimal,
  title={Optimal adaptive control for quantum metrology with time-dependent Hamiltonians},
  author={Pang, Shengshi and Jordan, Andrew N},
  journal={Nature communications},
  volume={8},
  number={1},
  pages={14695},
  year={2017},
  publisher={Nature Publishing Group UK London}
}

@article{PhysRevLett.117.160801,
  title = {Sequential Feedback Scheme Outperforms the Parallel Scheme for Hamiltonian Parameter Estimation},
  author = {Yuan, Haidong},
  journal = {Phys. Rev. Lett.},
  volume = {117},
  issue = {16},
  pages = {160801},
  numpages = {6},
  year = {2016},
  month = {Oct},
  publisher = {American Physical Society},
  doi = {10.1103/PhysRevLett.117.160801},
  url = {https://link.aps.org/doi/10.1103/PhysRevLett.117.160801}
}

@article{hu2024control,
  title={Control incompatibility in multiparameter quantum metrology},
  author={Hu, Zhiyao and Wang, Shilin and Qiao, Linmu and Isogawa, Takuya and Li, Changhao and Yang, Yu and Wang, Guoqing and Yuan, Haidong and Cappellaro, Paola},
  journal={arXiv preprint arXiv:2411.18896},
  year={2024}
}

@article{isogawa2025entanglement,
  title={Entanglement-assisted multiparameter estimation with a solid-state quantum sensor},
  author={Isogawa, Takuya and Wang, Guoqing and Li, Boning and Hu, Zhiyao and Nishimura, Shunsuke and Kanamoto, Ayumi and Yuan, Haidong and Cappellaro, Paola},
  journal={arXiv preprint arXiv:2505.14578},
  year={2025}
}

@article{xie2019optimal,
  title={Optimal control for multi-parameter quantum estimation with time-dependent Hamiltonians},
  author={Xie, Dong and Xu, Chunling},
  journal={Results in Physics},
  volume={15},
  pages={102620},
  year={2019},
  publisher={Elsevier}
}

@article{long2025optimal,
  title={Optimal multi-parameter quantum metrology for the frequencies of magnetic field},
  author={Long, Zhenhua and Pang, Shengshi},
  journal={Chinese Physics B},
  year={2025}
}

@article{khaneja2005optimal,
  title={Optimal control of coupled spin dynamics: design of NMR pulse sequences by gradient ascent algorithms},
  author={Khaneja, Navin and Reiss, Timo and Kehlet, Cindie and Schulte-Herbr{\"u}ggen, Thomas and Glaser, Steffen J},
  journal={Journal of magnetic resonance},
  volume={172},
  number={2},
  pages={296--305},
  year={2005},
  publisher={Elsevier}
}

@article{dolde2014high,
  title={High-fidelity spin entanglement using optimal control},
  author={Dolde, Florian and Bergholm, Ville and Wang, Ya and Jakobi, Ingmar and Naydenov, Boris and Pezzagna, S{\'e}bastien and Meijer, Jan and Jelezko, Fedor and Neumann, Philipp and Schulte-Herbr{\"u}ggen, Thomas and others},
  journal={Nature communications},
  volume={5},
  number={1},
  pages={3371},
  year={2014},
  publisher={Nature Publishing Group UK London}
}

@article{rong2015experimental,
  title={Experimental fault-tolerant universal quantum gates with solid-state spins under ambient conditions},
  author={Rong, Xing and Geng, Jianpei and Shi, Fazhan and Liu, Ying and Xu, Kebiao and Ma, Wenchao and Kong, Fei and Jiang, Zhen and Wu, Yang and Du, Jiangfeng},
  journal={Nature communications},
  volume={6},
  number={1},
  pages={8748},
  year={2015},
  publisher={Nature Publishing Group UK London}
}

@article{PhysRevA.100.012110,
  title = {Optimal quantum optical control of spin in diamond},
  author = {Tian, Jiazhao and Du, Tianyi and Liu, Yu and Liu, Haibin and Jin, Fangzhou and Said, Ressa S. and Cai, Jianming},
  journal = {Phys. Rev. A},
  volume = {100},
  issue = {1},
  pages = {012110},
  numpages = {12},
  year = {2019},
  month = {Jul},
  publisher = {American Physical Society},
  doi = {10.1103/PhysRevA.100.012110},
  url = {https://link.aps.org/doi/10.1103/PhysRevA.100.012110}
}

@article{liddy2023optimal,
  title={Optimal control theory techniques for nitrogen vacancy ensembles in single crystal diamond},
  author={Liddy, Madelaine SZ and Borneman, Troy and Sprenger, Peter and Cory, David},
  journal={Quantum Information Processing},
  volume={22},
  number={10},
  pages={358},
  year={2023},
  publisher={Springer}
}

@article{tovsner2009optimal,
  title={Optimal control in NMR spectroscopy: Numerical implementation in SIMPSON},
  author={To{\v{s}}ner, Zden{\v{e}}k and Vosegaard, Thomas and Kehlet, Cindie and Khaneja, Navin and Glaser, Steffen J and Nielsen, Niels Chr},
  journal={Journal of Magnetic Resonance},
  volume={197},
  number={2},
  pages={120--134},
  year={2009},
  publisher={Elsevier}
}

@article{de2011second,
  title={Second order gradient ascent pulse engineering},
  author={de Fouquieres, Pierre and Schirmer, Sophie G and Glaser, Steffen J and Kuprov, Ilya},
  journal={Journal of Magnetic Resonance},
  volume={212},
  number={2},
  pages={412--417},
  year={2011},
  publisher={Elsevier}
}

@article{PhysRevA.90.033628,
  title = {Optimal quantum control of Bose-Einstein condensates in magnetic microtraps: Comparison of gradient-ascent-pulse-engineering and Krotov optimization schemes},
  author = {J\"ager, Georg and Reich, Daniel M. and Goerz, Michael H. and Koch, Christiane P. and Hohenester, Ulrich},
  journal = {Phys. Rev. A},
  volume = {90},
  issue = {3},
  pages = {033628},
  numpages = {9},
  year = {2014},
  month = {Sep},
  publisher = {American Physical Society},
  doi = {10.1103/PhysRevA.90.033628},
  url = {https://link.aps.org/doi/10.1103/PhysRevA.90.033628}
}

@article{dionis2025optimal,
  title={Optimal control of a Bose-Einstein condensate in an optical lattice: the non-linear and two-dimensional cases},
  author={Dionis, Etienne and Peaudecerf, Bruno and Gu{\'e}rin, St{\'e}phane and Gu{\'e}ry-Odelin, David and Sugny, Dominique},
  journal={Frontiers in Quantum Science and Technology},
  volume={4},
  pages={1540695},
  year={2025},
  publisher={Frontiers Media SA}
}

@article{xu2019generalizable,
  title={Generalizable control for quantum parameter estimation through reinforcement learning},
  author={Xu, Han and Li, Junning and Liu, Liqiang and Wang, Yu and Yuan, Haidong and Wang, Xin},
  journal={npj Quantum Information},
  volume={5},
  number={1},
  pages={82},
  year={2019},
  publisher={Nature Publishing Group UK London}
}

@article{PhysRevA.103.042615,
  title = {Generalizable control for multiparameter quantum metrology},
  author = {Xu, Han and Wang, Lingna and Yuan, Haidong and Wang, Xin},
  journal = {Phys. Rev. A},
  volume = {103},
  issue = {4},
  pages = {042615},
  numpages = {13},
  year = {2021},
  month = {Apr},
  publisher = {American Physical Society},
  doi = {10.1103/PhysRevA.103.042615},
  url = {https://link.aps.org/doi/10.1103/PhysRevA.103.042615}
}

@article{PhysRevLett.115.110401,
  title = {Optimal Feedback Scheme and Universal Time Scaling for Hamiltonian Parameter Estimation},
  author = {Yuan, Haidong and Fung, Chi-Hang Fred},
  journal = {Phys. Rev. Lett.},
  volume = {115},
  issue = {11},
  pages = {110401},
  numpages = {7},
  year = {2015},
  month = {Sep},
  publisher = {American Physical Society},
  doi = {10.1103/PhysRevLett.115.110401},
  url = {https://link.aps.org/doi/10.1103/PhysRevLett.115.110401}
}

@article{PhysRevA.96.012117,
  title = {Quantum parameter estimation with optimal control},
  author = {Liu, Jing and Yuan, Haidong},
  journal = {Phys. Rev. A},
  volume = {96},
  issue = {1},
  pages = {012117},
  numpages = {14},
  year = {2017},
  month = {Jul},
  publisher = {American Physical Society},
  doi = {10.1103/PhysRevA.96.012117},
  url = {https://link.aps.org/doi/10.1103/PhysRevA.96.012117}
}

@article{PhysRevA.95.062342,
  title = {Enhancing sensitivity in quantum metrology by Hamiltonian extensions},
  author = {Fra\"{\i}sse, Julien Mathieu Elias and Braun, Daniel},
  journal = {Phys. Rev. A},
  volume = {95},
  issue = {6},
  pages = {062342},
  numpages = {10},
  year = {2017},
  month = {Jun},
  publisher = {American Physical Society},
  doi = {10.1103/PhysRevA.95.062342},
  url = {https://link.aps.org/doi/10.1103/PhysRevA.95.062342}
}

@article{PhysRevLett.123.250502,
  title = {Maximal Quantum Fisher Information for Mixed States},
  author = {Fiderer, Lukas J. and Fra\"{\i}sse, Julien M. E. and Braun, Daniel},
  journal = {Phys. Rev. Lett.},
  volume = {123},
  issue = {25},
  pages = {250502},
  numpages = {6},
  year = {2019},
  month = {Dec},
  publisher = {American Physical Society},
  doi = {10.1103/PhysRevLett.123.250502},
  url = {https://link.aps.org/doi/10.1103/PhysRevLett.123.250502}
}

@article{PhysRevLett.111.070403,
  title = {Quantum Enhanced Multiple Phase Estimation},
  author = {Humphreys, Peter C. and Barbieri, Marco and Datta, Animesh and Walmsley, Ian A.},
  journal = {Phys. Rev. Lett.},
  volume = {111},
  issue = {7},
  pages = {070403},
  numpages = {5},
  year = {2013},
  month = {Aug},
  publisher = {American Physical Society},
  doi = {10.1103/PhysRevLett.111.070403},
  url = {https://link.aps.org/doi/10.1103/PhysRevLett.111.070403}
}

@article{PhysRevResearch.4.043057,
  title = {QuanEstimation: An open-source toolkit for quantum parameter estimation},
  author = {Zhang, Mao and Yu, Huai-Ming and Yuan, Haidong and Wang, Xiaoguang and Demkowicz-Dobrza\ifmmode \acute{n}\else \'{n}\fi{}ski, Rafa\l{} and Liu, Jing},
  journal = {Phys. Rev. Res.},
  volume = {4},
  issue = {4},
  pages = {043057},
  numpages = {38},
  year = {2022},
  month = {Oct},
  publisher = {American Physical Society},
  doi = {10.1103/PhysRevResearch.4.043057},
  url = {https://link.aps.org/doi/10.1103/PhysRevResearch.4.043057}
}

\end{document}